\documentclass[aps,pra,twocolumn,showpacs,floatfix,superscriptaddress,reprint,
nobalancelastpage]{revtex4-1}
\usepackage[pdftex]{color}
\usepackage[english]{babel}
\usepackage{amsfonts}
\usepackage{amsmath}
\usepackage{graphicx}
\usepackage[caption=false]{subfig}
\usepackage[modulo]{lineno}

\usepackage{soul}
\usepackage{color}
\usepackage{ulem}

\newcommand{\si}[1]{#1}

\newcommand{\eps}{\varepsilon}
\newcommand{\veps}{\epsilon}
\newcommand{\expo}[1]{\mathrm{exp}\left(#1\right)}
\newcommand{\expoc}[1]{\mathrm{exp}\left[#1\right]}
\newcommand{\expol}[1]{\mathrm{exp}\left\{#1\right\}}
\newcommand{\cose}[1]{\mathrm{cos}\left(#1\right)}
\newcommand{\cosec}[1]{\mathrm{cos}\left[#1\right]}
\newcommand{\seno}[1]{\mathrm{sin}\left(#1\right)}
\newcommand{\senoc}[1]{\mathrm{sin}\left[#1\right]}

\newcommand{\Ket}[1]{\left|#1\right>}

\newcommand{\BraKet}[2]{\left<#1|#2\right>}

\begin{document}

\title{Maximum population transfer in a periodically driven quantum system}

\author{P. M. Poggi}
\email{ppoggi@df.uba.ar}
\affiliation{Departamento de F\'{\i}sica Juan Jose Giambiagi 
 and IFIBA CONICET-UBA,
 Facultad de Ciencias Exactas y Naturales, Ciudad Universitaria, 
 Pabell\'on 1, 1428 Buenos Aires, Argentina}
\affiliation{Grupo de Sistemas Complejos, 
 Escuela T\'ecnica Superior de Ingenieros Agr\'onomos, 
 Universidad Polit\'ecnica de Madrid, 28040 Madrid, Spain}
\affiliation{Instituto de Ciencias Matem\'aticas (ICMAT), 
 Universidad Aut\'onoma de Madrid, 
 Cantoblanco, 28049 Madrid, Spain}
\author{F. J. Arranz}
\affiliation{Grupo de Sistemas Complejos, 
 Escuela T\'ecnica Superior de Ingenieros Agr\'onomos, 
 Universidad Polit\'ecnica de Madrid, 28040 Madrid, Spain}
\author{R. M. Benito}
\affiliation{Grupo de Sistemas Complejos, 
 Escuela T\'ecnica Superior de Ingenieros Agr\'onomos, 
 Universidad Polit\'ecnica de Madrid, 28040 Madrid, Spain}
\author{F. Borondo}
\affiliation{Instituto de Ciencias Matem\'aticas (ICMAT), 
 Universidad Aut\'onoma de Madrid, 
 Cantoblanco, 28049 Madrid, Spain}
\affiliation{Departamento de Qu\'{\i}mica, 
 Universidad Aut\'onoma de Madrid, 
 Cantoblanco, 28049 Madrid, Spain}
\author{D. A. Wisniacki} 
\affiliation{Departamento de F\'{\i}sica Juan Jose Giambiagi 
 and IFIBA CONICET-UBA,
 Facultad de Ciencias Exactas y Naturales, Ciudad Universitaria, 
 Pabell\'on 1, 1428 Buenos Aires, Argentina}
\date{\today}

\begin{abstract}

We study the dynamics of a two-level quantum system under the influence of sinusoidal driving in the intermediate frequency regime. Analyzing the Floquet quasienergy spectrum, we find combinations of the field parameters for which population transfer is optimal and takes place through a series of well defined steps of fixed duration. We also show how the corresponding evolution operator can be approximated 
at all times by a very simple analytical expression. \si{We propose this model as being specially suitable for treating periodic driving at avoided crossings found in complex multi-level systems, and thus show a relevant application of our results to designing a control protocol in a realistic molecular model}.
\end{abstract}

\maketitle

\section{Introduction}

Understanding the coherent \si{manipulation} of quantum systems using time-dependent 
interacting fields is a goal of primary interest in many different areas, 
including chemical reactivity \cite{bib:chem}, nanotechnology \cite{bib:nano}, 
and quantum information processing \cite{bib:qc}. 
To this end, simple analytically solvable two-level systems (TLS) 
are often used since they can efficiently describe the dynamics. 
One popular choice is the Landau-Zener model \cite{bib:zener}, 
in which the driving field is assumed to vary linearly with time. 
Nonetheless, in many experimental situations sinusoidal, 
time-periodic control fields are easier to produce and manipulate, 
and are thus the preferred option. 
Beyond the well-known Rabi model (which accounts for the weak driving case), 
many approaches have been used in the literature to describe various non-trivial limits of this type of systems \cite{bib:floquet,bib:nori,bib:wu,bib:dassarma}.\\

A striking phenomenon induced by time-periodic fields is 
the so-called coherent destruction of tunneling (CDT), 
first predicted by Grossmann \textit{et al.}~\cite{bib:grossman} 
and then observed experimentally \cite{bib:cdtexp1}. 
A particle in a symmetric double-well potential usually oscillates back and forth, 
if initially localized in one of the wells. 
However, if the depth of the wells oscillates in time, 
the tunneling rate may dramatically change. 
Actually, for certain combinations of the driving parameters, 
the rate vanishes, resulting in an effective localization of 
the particle in the initial well. 
As previously shown \cite{bib:grossman, bib:deg}, this behavior takes 
place only when some Floquet quasienergies are degenerate.\\

In this work, we show that the Floquet spectrum of a TLS under 
a sinusoidal driving in the regime of intermediate frequencies 
($\omega\simeq\Delta$, being $\omega$ the driving frequency and $\Delta$ the characteristic frequency of the system \cite{bib:hbarra}) \si{shows} a second kind of 
``special points'', defined by the condition that the quasienergy 
separation is a local maximum, where: 
(i) population inversion is achieved after a time interval that only 
depends on the quasienergy difference; 
(ii) the evolution of the populations 
happens through a series of well-defined steps of fixed duration, 
in which the probability remains approximately constant, and 
(iii) the full time-dependent evolution operator $U(t)$ can be 
obtained in a very simple analytical form, which provides a clear 
physical interpretation of (ii).

\si{Finally, taking into account the general validity of this two-level model, we study at what extent the results we obtain can be applied to multi-level systems which are periodically driven at an avoided crossings (AC). By designing of a control protocol in a realistic model for the LiNC$\rightleftharpoons$LiCN isomerization} \cite{bib:brocks,bib:murgida}, \si{we find that the intermediate frequency regime is specially suitable for such complex systems}.\\

This paper is organized as follows. In Sec. II we present the model system, enumerate the main results of the well-known high frequency regime and present the basics of the Floquet formalism. We then turn to the intermediate frequency regime, where we show that the dynamics of the system changes considerably, showing a remarkably regular behavior for certain values of the driving field amplitude. In Sec. III we ellaborate on the analysis and interpretation of these results, and develop a very simple Bloch sphere model wich allows us to get an analytical solution for the evolution operator. Finally, in Sec. IV we describe the LiCN/LiNC molecular system and propose a control protocol suitable for achieving an isomerization reaction. Sec. V contains some concluding remarks.

\section{Periodically driven two-level systems}

We consider a hamiltonian of the form
\begin{equation}
H(t) = \frac{\Delta}{2}\sigma_x + \eps(t)\sigma_z,
  \label{ec:hami}
\end{equation}
where $\sigma_x$ and $\sigma_z$ are the usual Pauli operators. 
The instantaneous eigenvalues of $H$ as a function of the control 
parameter $\eps$ show the usual avoided crossing (AC) structure, 
reaching a minimal separation of $\Delta$ at $\eps=0$. 
We consider the driving field to be $\eps(t)=A\:\cose{\omega t}$, 
and define $T=2\pi/\omega$ as the period of $H(t)$. 
When dealing with this type of systems, it is customary to factorize 
the evolution operator as $U(t)=U_1(t)\: U_2(t)$, 
where $U_1(t)=\expoc{-i\:\gamma_z(t)\:\sigma_z/2}$ 
can be regarded as a transformation to a rotating frame, 
since $\gamma_z(t)=2\int\eps(\tau)\:d\tau=(2A/\omega)\seno{\omega t}$. 
The remaining factor is obtained by the transformed Schr\"odinger 
equation $i\dot{U_2}(t)=H_2(t)\:U_2(t)$, 
where $H_2(t) = U_1^{\dagger}HU_1$ as
%
\begin{equation}
   H_2(t) =\frac{\Delta}{2}\left\{\cosec{\gamma_z(t)}\sigma_x
          +\senoc{\gamma_z(t)}\sigma_y\right\}
  \label{ec:hamief}
\end{equation}
The time dependence in this expression can be averaged out over one 
period of the driving field in the high frequency regime, 
i.e.~$\omega\gg\Delta$, using the rotating wave approximation (RWA) \cite{bib:nori,bib:deg}. 
This gives $U_2(t)=\expo{-i\frac{\Delta'}{2}t\:\sigma_x}$, 
with $\Delta'=\Delta J_0\left(2A/\omega\right)$, 
being $J_0$ a Bessel function. 
For the values of $2A/\omega$ corresponding to the zeros of $J_0$, 
the evolution operator $U(t)$ is diagonal in the $\sigma_z$ basis set, 
$\left\{\Ket{0},\Ket{1}\right\}$, which explains the occurrence 
of the CDT phenomenon. 
For any other value of the amplitude $A$ the population inversion 
between these states takes place in a finite lapse of time, given by
\begin{equation}
  T_F = \frac{\pi}{\Delta'}.\\
  \label{ec:tf}
\end{equation}
\begin{figure}[!tb]
\includegraphics[width=\linewidth]{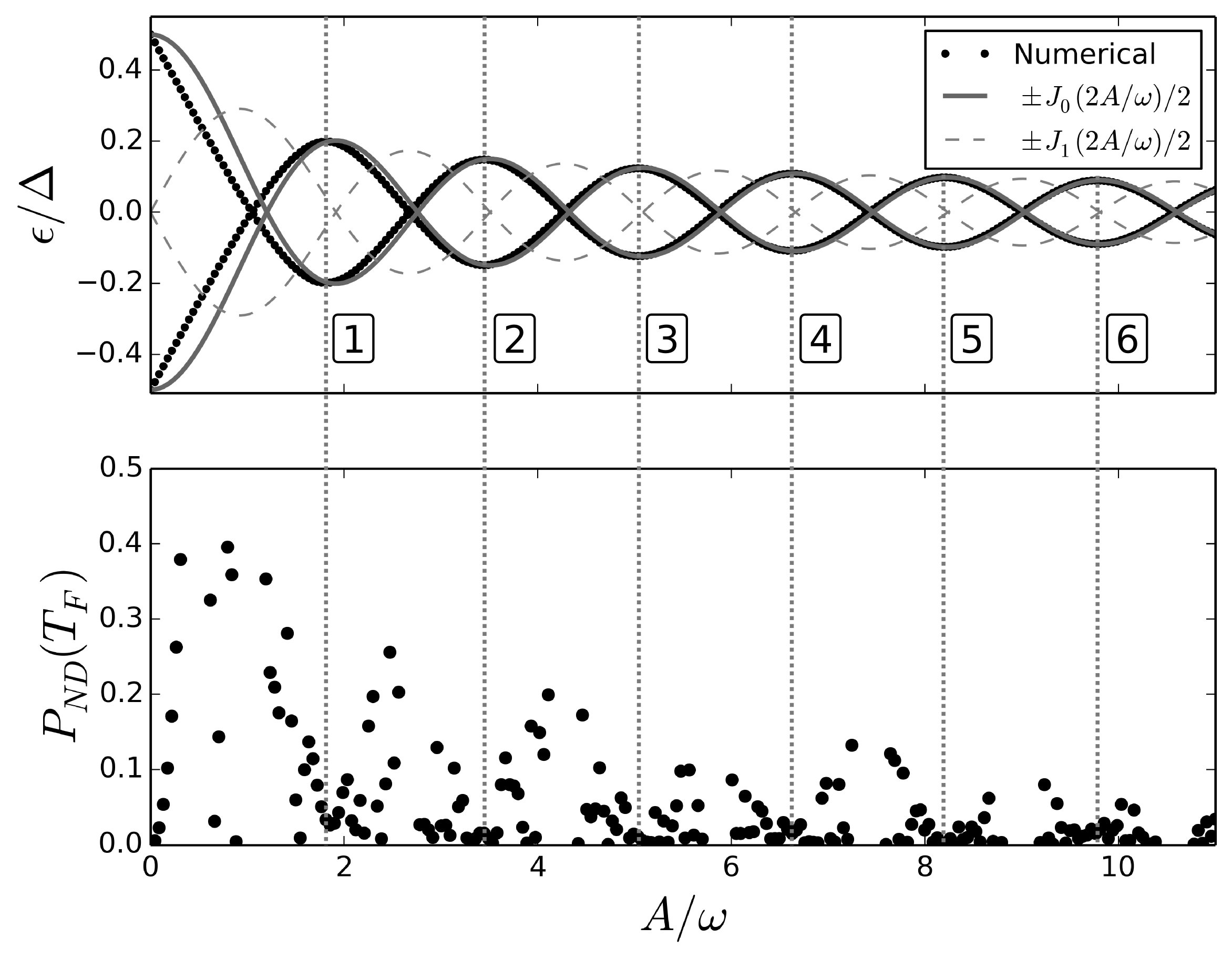}
\caption{\label{fig:fig1} Top: Quasinergy spectrum for the two level hamiltonian of eq. (\ref{ec:hami}) as a function of $A/\omega$. The boxed numbers $n=1,2,\ldots$ label the points of the spectrum in which the quasienergy separation is locally maximal. On top, we plot the analytical expression (\ref{ec:cuasie}). Bottom: Non decay probability $P_{ND}$ at time $T_F$, defined by eq. (\ref{ec:tf}), calculated by numerical simulations of the system prepared in the initial state $\Ket{0}$, as a function of $A/\omega$. Near the degeneracies, where $T_F$ diverges, results are not displayed. In all cases, the resonant case  (i.e. $\omega=\Delta$) is studied.}
\end{figure}

When the RWA cannot be applied, a more general framework has to be used. 
In this case, Floquet theory \cite{bib:floquet} shows that for a time-periodic hamiltonian a full set of orthonormal solutions for the corresponding 
Schr\"odinger equation exists, which are of the form 
$\Ket{\Psi_\alpha(t)}=\expo{-i\epsilon_\alpha t}\Ket{\Phi_\alpha(t)}$, 
with $\alpha=0,1$ for a TLS.
The real-valued quantities $\left\{\epsilon_\alpha\right\}$ are called \textit{quasienergies}, and the states $\left\{\Ket{\Phi_\alpha(t)}\right\}$, 
which share the periodicity of $H(t)$, are called \textit{Floquet states}. 
The quasienergies can be obtained in an easy way by diagonalizing $U(T)$, 
something that can be done by numerically computing the time evolution 
from $t=0$ to $t=T$ of an adequate basis set. 
In this way, the eigenphases of $U(T)$ give the desired quasienergies, 
which in the case $\omega\gg\Delta$, discussed above, 
simply correspond to 
%
\begin{equation}
  \veps_\pm = \pm\frac{\Delta'}{2}.
  \label{ec:cuasie}
\end{equation}
This expression implies that the spectrum contains an infinite set 
of degeneracies as $A/\omega$ increases, and also that
expression (\ref{ec:tf}) can be rewritten as $T_F = \pi/|\veps_+-\veps_-|$.\\

When computed for lower frequencies, the quasienergy spectrum 
changes considerably for small amplitudes \cite{bib:creffield}, 
as shown in Fig.~\ref{fig:fig1} (top) for the case $\omega=\Delta$. 
However, the results still show the typical ribbon structure \cite{bib:ribbon}, 
and expression (\ref{ec:cuasie}) remains a reasonable approximation for 
$A/\omega\gtrsim 3$. \si{In order to compare the population dynamics in this model as opposed to the high-frequency regime, we study the validity of expression (\ref{ec:tf}). For that purpose}, we simulate the evolution of the system starting from $\Ket{0}$ 
for different values of amplitude, calculating the non-decay 
probability $P_{ND}(t)=|\BraKet{0}{\psi(t)}|^2$ at time $t=T_F$, 
in each case. The results that are shown in Fig.~\ref{fig:fig1} (bottom) reflect 
a more complex behavior than that predicted by the high frequency model, 
in which $P_{ND}(T_F)=0$ is expected for every value of $A$ corresponding 
to finite $T_F$.\\

\begin{figure}[!tb]
\includegraphics[width=\linewidth]{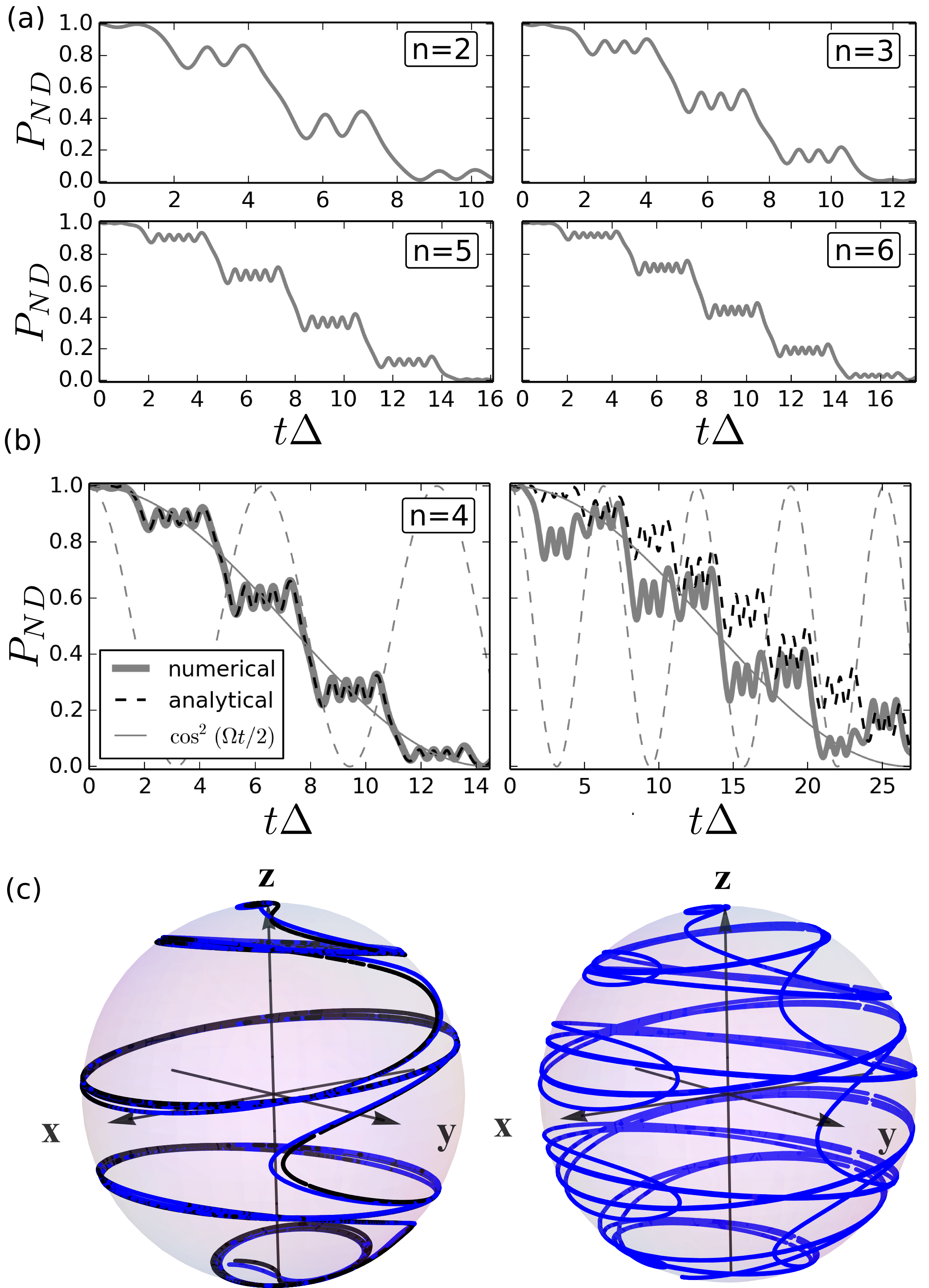}
\caption{\label{fig:fig2} 
(Color online) Time evolution of the non decay probability for the system starting in state $\Ket{0}$, 
in the resonant ($\omega=\Delta$) regime. 
(a) Amplitudes corresponding to peaks $n=2,3,5,6$. 
(b) Left: Amplitude corresponding to $n=4$. 
    Right: $A=4.5\omega$, between $n=2$ and $n=3$. 
Thick lines show the results given by numerical simulations, 
while the black dashed curve is given by the analytical solution 
(see text for details). 
For comparison, we show the solution predicted in the high frequency regime 
(solid light gray line), and a cosine function with the frequency of the driving field, 
$\omega$ (dashed light gray line). (c) Time evolution of the complete state of the system, depicted in Bloch sphere, for the cases shown in (b). The analytical solution is not shown for the case $A=4.5\omega$, for clarity.}
\end{figure}

More interesting is the fact that the results of Fig.~\ref{fig:fig1}
reveal the existence of a new outstanding feature: 
the points for which $P_{ND}\simeq 0$ pack around certain values of $A$, 
which correspond to the points of local maximum separation between  
quasienergies, i.e.~the ``peaks'' of the spectrum. 
We have labeled these points by $n=1,2,\ldots$ in the figure. 
To analyze this behavior in more detail, 
we consider the time evolution of $P_{ND}(t)$ for different values 
of the driving amplitude. 
Some representative numerical results are shown in Fig.~\ref{fig:fig2},
where it can be seen that $P_{ND}$ shows a ``ladder''-type structure, 
decreasing through a series of steps, in each of which the probability 
oscillates rapidly around a constant mean value. 
Moreover, as $n$ grows, the frequency of these oscillations increases, 
while the corresponding amplitude decreases. 
These steps occur whenever the field $\eps(t)$ reaches a maximum 
or a minimum, and then their amount can be estimated by the ratio 
$2\omega/\Omega$, with $\Omega=2\pi/T_F$. \si{We point out that the ocurrence of stepwise population inversion has been reported previously in this model \cite{bib:nori,bib:vitanov}, and can be accounted for using the transfer matrix approach in the limit of large amplitudes ($A/\omega\gg 1$). Here, we are interested in analyzing the particular conditions under which this behaviour takes place,} \si{specially because} when $A$ is set outside the peaks, the rapid oscillations still take place, but the ``stairs'' become worse defined, and the probability ladder may not necessarily be decreasing at all times, as illustrated in Fig.~\ref{fig:fig2} (b). 

\section{Maximum population transfer: Bloch sphere model and analytical solution}
\begin{figure}[!tb]
\includegraphics[width=0.95\linewidth]{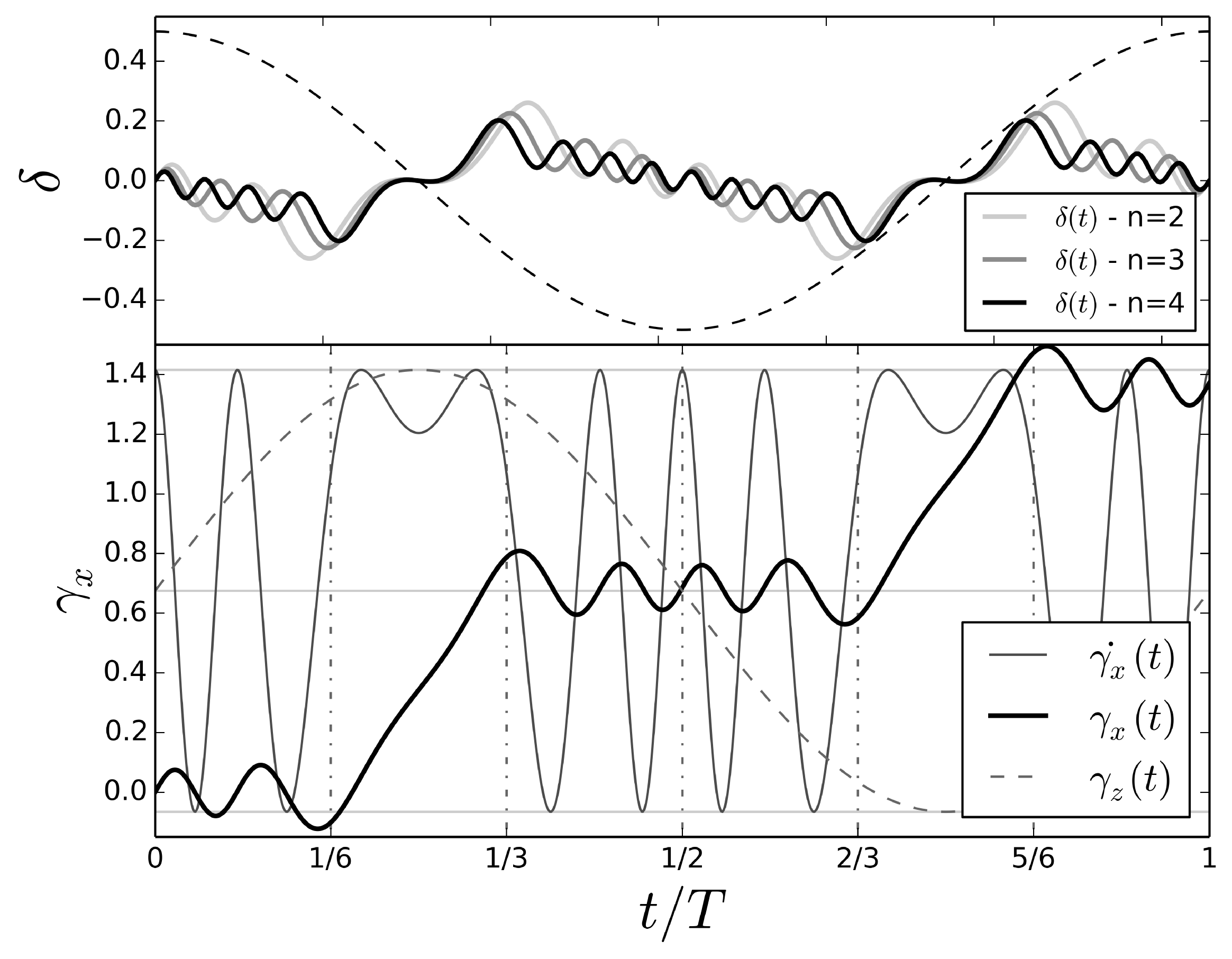}
\caption{\label{fig:fig3} 
Top: Plots of function $\delta(t)$ defined by Eq.~(\ref{ec:delta}), 
over one period of the driving field, for different values of amplitude 
corresponding to $n=2,3,4$. 
In dashed lines a cosine function with the frequency of the driving 
field $\omega$ is shown. 
Bottom: Plot of the time-dependent phases $\gamma_x(t)$ and $\gamma_z(t)$ 
which appear in the analytical solution $U(t)$ proposed for the evolution operator. 
Also shown is $B_x(t)=\dot{\gamma_x}(t)$. Note that the left axis labels correspond only to $\gamma_x(t)$, the remaining functions being properly normalized for comparision.}
\end{figure}

The singular behavior shown by the dynamics at the extrema of the quasienergy 
spectrum admits a (deeper) analytical explanation.
Hamiltonian $H_2$ in eq.~(\ref{ec:hamief}) can be regarded as equivalent
to the interaction of a spin-$1/2$ particle with a unit intensity magnetic field
$\vec{B(t)}$ rotating periodically but non-uniformly in the $x-y$ plane, 
such that the instantaneous Larmor frequency is $\Delta$. 
The components of this field can be expanded in Fourier series
%
 \begin{eqnarray}
  B_x(t)&\equiv&\cose{\gamma_z}=J_0\left(\nu\right) + 
    2\sum_{n=1}^{\infty} J_{2n}\left(\nu\right)\cose{2n\omega t} \label{ec:fourierc} \\
  B_y(t)&\equiv&\seno{\gamma_z}=2\sum_{n=1}^{\infty} J_{2n-1}\left(\nu \right)
           \senoc{(2n-1)\omega t}, \label{ec:fouriers}
\end{eqnarray}
where $\nu =2A/\omega$. 
If considered separately, the time integrals of both components give the 
accumulated phase throughout the evolution. 
As shown in Fig.~\ref{fig:fig3}, the contribution $\gamma_x(t)\equiv\int_0^tB_x(s)ds$ 
shows the ladder structure found previously, as the result of integrating 
a constant term added to an oscillating series. 
On the other hand, integrating $B_y(t)$ shows that the leading term vanishes 
when $J_1(2A/\omega)=0$, this resulting in a small phase contribution 
of the whole series. 
Also notice that, because of the relation $J'_0(x)=-J_1(x)$, 
the zeros of $J_1$ match the extrema of $J_0$, also giving the position 
of the spectrum peaks mentioned above, 
as long as approximation (\ref{ec:cuasie}) holds. 
In this situation, $U_2(t)$ is well approximated by 
$U_2(t)=\expoc{-\frac{i}{2}\gamma_x(t)\sigma_x}=
\expol{-\frac{i}{2}\left[\Delta't+\delta(t)\right]\sigma_x}$ with
%
\begin{equation}
 \delta(t)=\frac{\Delta}{\omega}\sum_n \frac{J_{2n}(2A/\omega)}{n}\:
           \seno{2n\omega t},
 \label{ec:delta}
\end{equation}
which is $T$-periodic and can be seen to vanish in the limit $\Delta/\omega\rightarrow 0$, 
as expected. \\

Plots of $\delta(t)$ for different values of $n$ are displayed 
in Fig.~\ref{fig:fig3}. 
This model approximates very well the population dynamics when the field 
parameters are set at the extrema of the quasienergy spectrum. 
A representative example is shown in Fig.~\ref{fig:fig2}. In this case the full evolution operator becomes
\begin{equation}
  U=U_1U_2=\expoc{-i\gamma_z(t)\sigma_z/2}\expoc{-i\gamma_x(t)\sigma_x/2},
  \label{ec:opevol}
\end{equation}
and the particular time-dependence of $\gamma_x$ and $\gamma_z$ over 
one period of the driving field (see Fig.~\ref{fig:fig3}) allows to 
rationalize the resulting dynamics, as follows. 
Let us consider a partition of the driving period in six equal intervals, 
each one of length $T/6$. 
Then $\gamma_x(t)$ and $\gamma_z(t)$ can be approximated as a sequence of linear 
and constant pieces, both showing opposite behaviors during the interval. 
That is, from $t=T/6$ to $t=2T/6$, $\gamma_z$ is almost constant and $\gamma_x$ 
increases with a positive slope; the resulting $U(t)$ being then well 
approximated by a $x$-rotation in Bloch sphere. 
From $t=2T/6$ to $t=3T/6$ (and also in the following interval)
$\gamma_x$ shows low-amplitude oscillations around a steady value, 
while $\gamma_z$ decreases in time almost linearly;
$U(t)$ will then produce rapid rotations around the $z$-axis 
rendering nearly constant populations.
Similarly, we can continue with the rest of the intervals in the period.
Finally, note that this discussion also accounts for the phenomenon of 
optimal population transfer at these points, 
shown in Fig.~\ref{fig:fig1}. 
Using this model, a simple calculation gives 
$P_{ND}(T_F)=\mathrm{sin}^2\left(\delta(T_F)/2\right)$, 
which is numerically seen never exceeding $10^{-2}$. 

\section{An example: control of isomerization reactions}
\si{Let us discuss next the application of our results in a molecular control problem \cite{bib:int_prac}.} For this purpose, we consider the LiNC/LiCN molecular system 
that has been extensively studied in connection with the theoretical 
issue of quantum chaos \cite{bib:LiCN}, and also in the simulation of the  
LiNC$\rightleftharpoons$LiCN isomerization reaction \cite{bib:Hase}
in solution, where it was proven to provide the first unambiguous 
example of the elusive Kramers turnover \cite{bib:Muller}. In general, isomerization reactions have generated a lot interest from the 
theoretical side \cite{bib:Hase} and also for their practical 
importance in many relevant chemical processes, specially of 
biological interest \cite{iso1,iso2,iso3,iso4}. For example, the control of the HCN isomerization was thoroughly 
studied in Refs.~\cite{control_HCN_1,control_HCN_2,control_HCN_3},
and the importance of intermediate states with configurations 
far from the usual ones discussed.

\subsection{The LiNC/LiCN molecular system}

The LiNC/LiCN isomerizing system presents two stable isomers at the 
linear configurations: Li--N--C and Li--C--N, which are separated by 
a relatively modest energy barrier of only 0.0157376 a.u. The C and N atoms are strongly bounded by a triple covalent bond, while the Li is attached to the CN moiety by mostly ionic forces,
due to the large charge separation existing between them.
For these reasons, the CN vibrational mode effectively decouples from the
other degrees of freedom of the molecule, and it can be considered frozen
at its equilibrium value, $r_e=2.186$.
On the other hand, the relative position of Li with respect to the center
of mass of the CN is much more flexible.
In particular the bending along the angular coordinate is very floppy,
and the corresponding vibration performs very large amplitude motions
even at moderate values of the excitation energy.
Accordingly, the vibrations of the whole system can be adequately described
by the following 2 degrees of freedom.
Using scattering or Jacobi coordinates $(R,r,\theta)$, where $R$ is the
distance from the Li atom to the center of mass of the CN fragment,
$r$ the C--N distance, and $\theta$ the angle formed by these two vectors,
the corresponding classical ($J=0$) Hamiltonian is given by
%
\begin{equation}
  H  =  \frac{P_R^2}{2 \mu_\mathrm{Li-CN}}
       + \frac{1}{2} \left( \frac{1}{\mu_\mathrm{Li-CN} R^2}+
            \frac{1}{\mu_\mathrm{CN} r_e^2} \right) P_{\theta}^2
          + V(R,\theta),
 \label{eq:H}
\end{equation}
where $P_R$ and $P_\theta$ are the associate conjugate momenta,
and the corresponding reduced masses are given by
$\mu_\mathrm{Li-CN}=m_\mathrm{Li} m_\mathrm{CN}/(m_\mathrm{Li}+m_\mathrm{CN})=10072$ and
$\mu_\mathrm{CN}=m_\mathrm{C} m_\mathrm{N}/m_\mathrm{CN}=11780$.\\

\begin{figure*}
\includegraphics[width=0.95\linewidth]{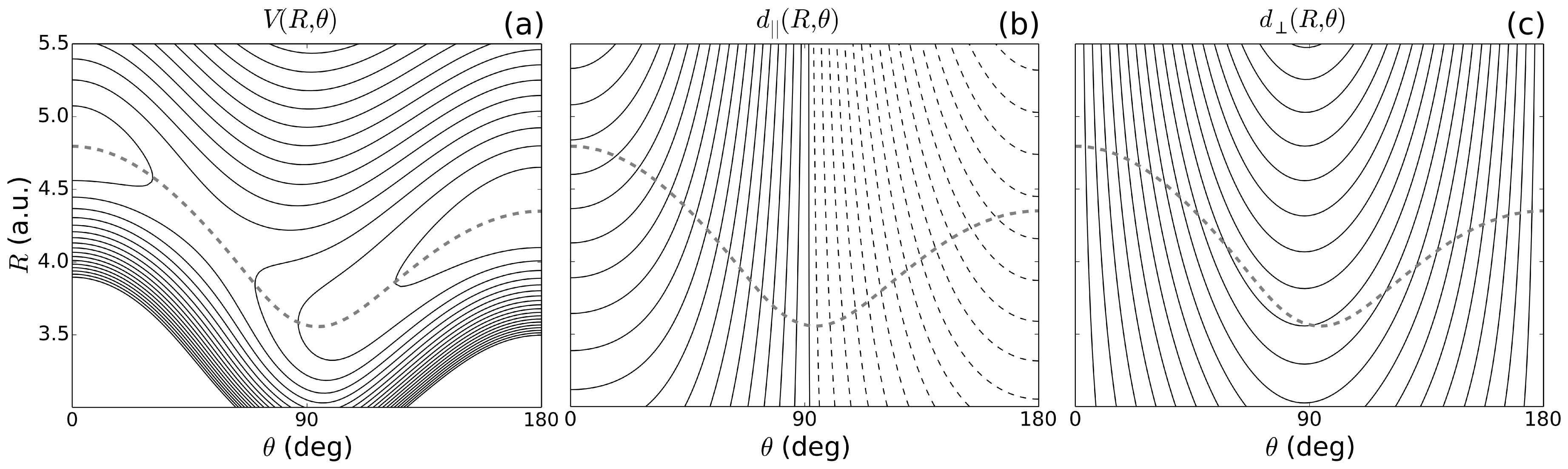}
\caption{Contour plot representation in the configuration space $(R,\theta)$ 
for the following functions of the LiNC/LiCN isomerizing system: 
(a) Potential energy surface $V(R,\theta)$ (contours every 1000 cm$^{-1}$ 
starting from zero). 
(b) Dipole moment component parallel to the N--C bond  
(contours every 0.25 a.u. starting from zero, dashed lines correspond to
negative values).
(c) Same as (b) for the perpendicular component.
The minimum energy path connecting the isomer wells has been 
superimposed as a grey thick dashed line in the three plots.}
  \label{fig:LiCN}
\end{figure*}

 Note that we assume that the isomerization process is fast compared 
with the rotation of the molecule. The potential interaction, $V(R,\theta)$, is given by a 10--terms expansion
in Legendre polynomials,
%
\begin{equation}
  V(R,\theta)= \sum_{\lambda=0}^9 v_\lambda(R) P_\lambda(\cos \theta),
\end{equation}
where the coefficients, $v_\lambda(R)$, are combinations of long and
short--term interactions whose actual expressions have been taken from
the literature \cite{LiCN.PES}.
This potential, which is shown in Fig.~\ref{fig:LiCN} (a) as a contour plot,
has a global minimum at $(R,\theta)=(4.349,\pi)$,
a relative minimum at $(R,\theta)=(4.795,0)$,
and a saddle point at $(R,\theta)=(4.221,0.292 \pi)$.
The two minima correspond to the stable isomers at the linear configurations,
LiNC and LiCN, respectively.
The LiNC configuration, $\theta=\pi$, is more stable than that for LiCN,
$\theta=0$.
The minimum energy path connecting the two isomers has also been plotted
superimposed in the figure with dashed line.\\

The LiCN molecule is a polar molecule, i.e., it has a permanent dipole moment, 
so that in the presence of an electric field, $\vec{{\cal E}}$, 
an additional potential energy term appears, this leading to the 
following effective Hamiltonian function
\begin{equation}
  H = H_{\mathrm{LiCN}} - \vec{d}(R,\theta)\cdot\vec{{\cal E}},
\label{ec:hamim}
\end{equation}   
where $\vec{d}(R,\theta)$ is the dipole moment of the LiNC/LiCN molecular system. 
For the dipole moment, we have taken from the literature the {\it ab initio} 
calculations fitted to an analytic expansion in associated Legendre functions 
of Brocks \textit{et al.} \cite{bib:brocks}.  
The corresponding components parallel and perpendicular to the N--C 
bond are shown in Fig.~\ref{fig:LiCN} (b) and (c), respectively.\\



%

\begin{figure}[!tb]
\includegraphics[width=\linewidth]{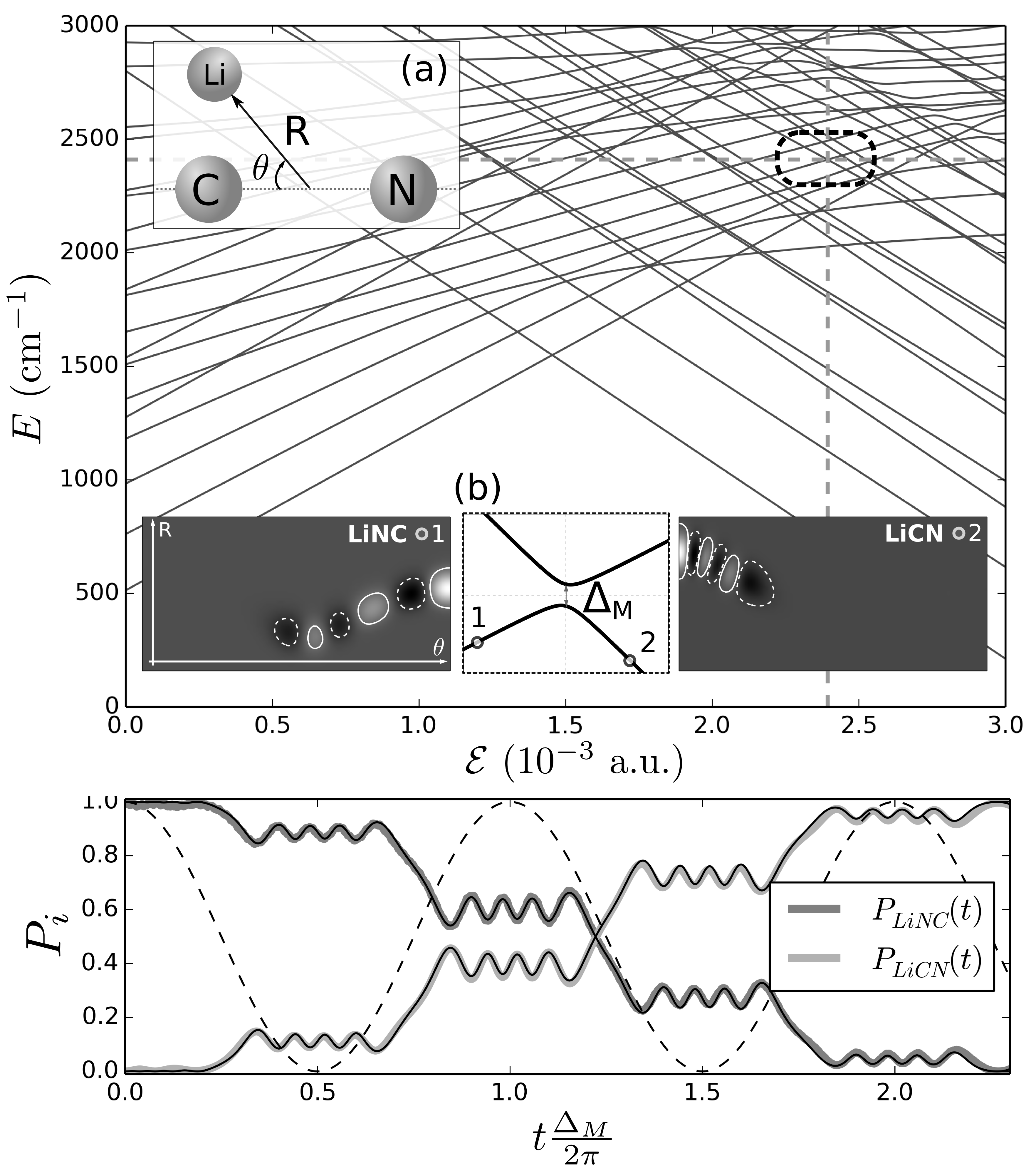}
\caption{\label{fig:fig5} 
Top: (main) energy spectrum of the LiNC/LiCN molecular system as a function 
of the electric field intensity $\cal{E}$. 
(a) Schematic diagram of the LiCN molecule, including the set of coordinates 
$(R,\theta)$ used. 
(b) Enlarged view of the squared zone of the spectrum, 
showing the AC considered in the control strategy. 
At both sides, density plots of the wavefunctions far from the AC, 
which represent excited isomerized states. The axis ranges are $0<\theta<180^\circ$ 
and $3<R<5.5$ a.u. 
Bottom: numerically simulated population evolution, starting in state 
$\Ket{1}$ and setting $\omega=\Delta_M$ and $A=(1.23\times10^{-2})\Delta$. 
Full black lines show the result predicted by the two-level analytical solution 
(\ref{ec:opevol}) applied to this system.}
\end{figure}
\subsection{\si{Achieving isomerization via a DC+AC field}}
In order to design an effective control strategy, we consider the electric field to be parallel to the C$\rightarrow$N bond and compute the vibrational level spectrum as a function of the (static) electric 
field intensity $\cal{E}$, as previously proposed in Refs. \cite{bib:murgida,bib:poggi}. In order to do so, we used the discrete variable representation - distributed Gaussian basis (DVR-DGB) method introduced in Ref.~\cite{bib:bacic}. Results are shown in Fig.~\ref{fig:fig5}.
As a rule of thumb, positive-slope energy lines correspond to LiNC states 
(that is, those localized in the $\theta=180^\circ$ well), 
and the negative-slope lines to LiCN states ($\theta\sim 0$). Further details on the structure of this spectrum can be found on Refs. \cite{Arranz1,Arranz2}.\\

A careful analysis of the spectrum shows that most ACs 
in the low-energy region are very narrow and thus correspond to 
interactions too weak to be useful in the control process. 
However, there is an AC centered at $\mathcal{E}=\mathcal{E}_{\rm{dc}}\equiv2.39\times10^{-3}$ a.u., 
with a spectral gap of $\Delta_M=0.15$ cm$^{-1}$ which seems suitable for
our purposes. 
Indeed, far for from the AC, the involved eigenstates, 
termed $\Ket{\mathrm{LiNC}}$ and $\Ket{\mathrm{LiCN}}$, show localization in opposite wells (see Fig.~\ref{fig:fig5}-b) as needed in the control process. 
We thus \si{analyze} the use of an electric field of the form: 
$\mathcal{E}(t)=\mathcal{E}_{\mathrm{dc}} + A\cose{\omega t}$.\\

\begin{figure}[!tb]
\includegraphics[width=\linewidth]{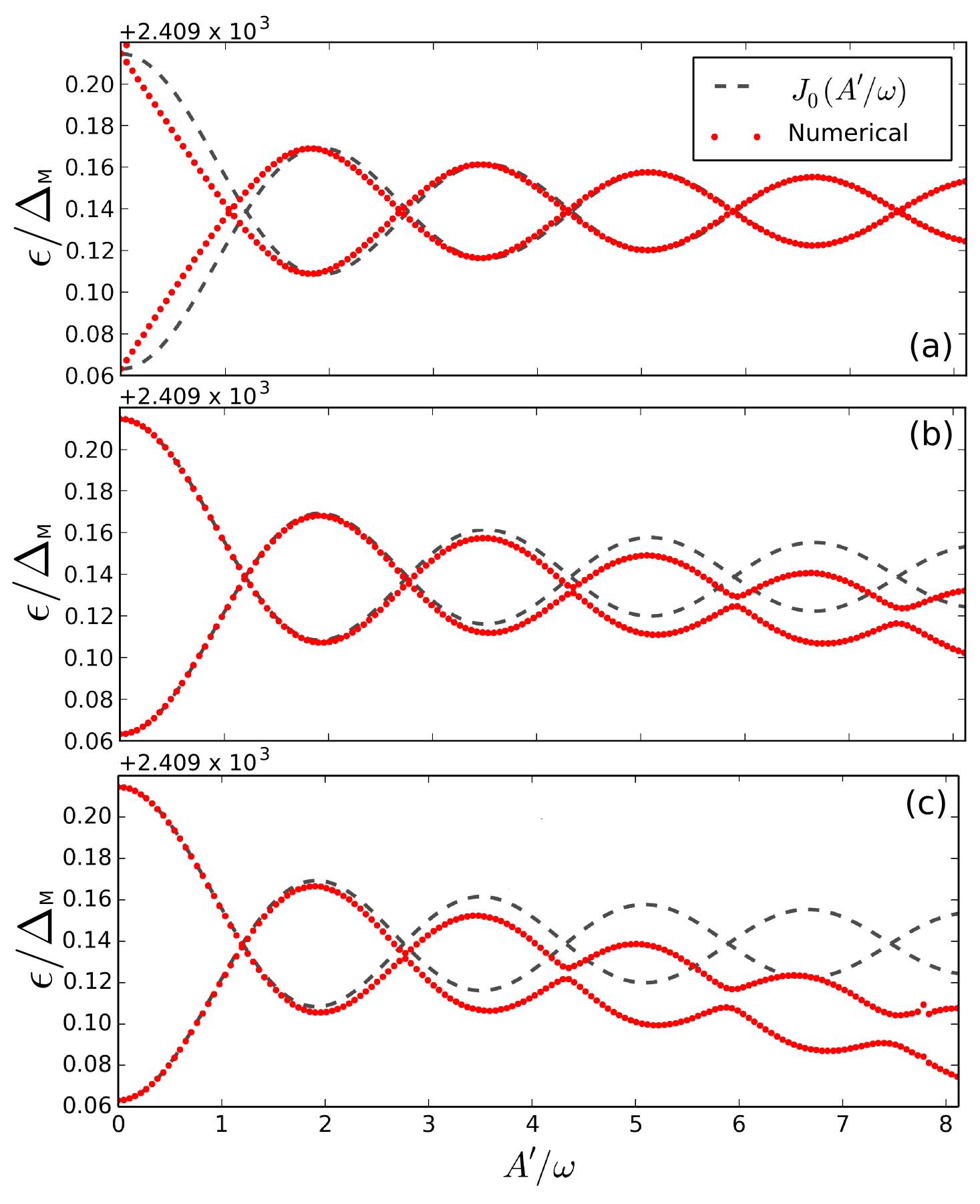}
\caption{(Color online) Quasienergy spectrum for the LiNC/LiCN molecular system, 
focused in the energy range of the avoided crossing under study
(see Fig.~5). 
The values of driving field frequency used in each case are: 
(a) $\omega=\Delta_M$, (b) $\omega=20\Delta_M$, (c) $\omega=30\Delta_M$.}
  \label{fig:fig6}
\end{figure}

\si{The main feature to be emphasized here is that the results drawn from usual high driving frequency regime could not be applied in this case.} This is because the quasienergy spectrum is a function of the ratio $A/\omega$, which means that high values of $\omega$ would then imply the use of large amplitudes. Note that, in such case, the control parameter (\si{i.e., $\mathcal{E}$ in our case}) would reach zones of the energy spectrum 
where more levels are involved in an effective interaction. This would invalidate the two-level approximation in a many level system showing multiple ACs, like ours. This can be clearly seen in Fig. \ref{fig:fig6}, where we show the quasienergy spectrum of our 
molecular system as a function of $A'/{\omega}$  \cite{bib:aclaracion}, for different 
values of driving frequency $\omega$. Note that only the region of the spectrum corresponding to the marked AC in the Fig.~4 (b) of the main text is displayed here. 
As can be seen, the ribbon structure typical of the kind of
system considered in this work becomes clearly distorted 
as the frequency increases. Therefore, we propose to work in the intermediate frequency regime discussed before,
so that the main results of the previous sections become relevant for this problem.
Actually, setting $\omega$ equal to $\Delta_M$ make expressions (\ref{ec:hami}) to (\ref{ec:opevol}) straightforwardly applicable.
As an illustration, we show in the bottom panel of Fig.~\ref{fig:fig5} 
the evolution obtained starting from state $\Ket{LiNC}$ \cite{bib:otromet} for
$A=(1.23\times10^{-2})\Delta$ (corresponding to $n=4$).
\si{In terms of control efficiency and suitability, we point out that the total control time is approximately $2.3\times 2\pi/\Delta_M\simeq0.51$ ps, which is well below the $400$ ps reported in Ref.~\cite{bib:murgida}. Nevertheless, it should be noted that this protocol would require fine tuning of the control parameters, similarly to more elemental strategies (such as applying a single pi-pulse}.  We remark that the results predicted by the analytical model proposed in Sec. IV are in full agreement with the numerical results, as can be seen in Fig. \ref{fig:fig5}.\\

\section{Final remarks} 

In summary, we have shown the existence of a set of special 
points in the quasienergy spectrum of a periodically-driven TLS, 
where the evolution of the populations takes place with maximum 
probability transfer. 
These points correspond to the maxima \si{and minima} in the typical ribbon 
structure exhibited by the spectrum, localized between 
the degeneracies predicted by the occurrence of CDT. 
We have also shown that for these particular combinations of 
the driving parameters the full evolution operator for the system 
can be well approximated by a very simple analytical expression, 
which reveals that the system evolves in a Bloch sphere following 
a sequence of rotations around the $x$ and $z$ axes. 
This behavior reflects in the appearance of a series of steps 
in the time evolution of the populations, 
whose average takes the form of a decreasing ``ladder'', \si{a behaviour which has been reported in previous works on this model} \cite{bib:nori,bib:vitanov} 
It should be noted that these results correspond to 
the intermediate frequency regime ($\omega\simeq\Delta$) 
where the RWA does not apply. 
Finally, we have made use of these conclusions
\si{to study the isomerization process induced by an oscillating electric field applied to a triatomic molecule}. 
Using this realistic model, we have shown that the regime 
described in this Letter is particularly relevant 
in many level systems showing multiple ACs, where the use of 
large-amplitude driving fields would make the simple two-level 
approximation invalid.
We also believe that the results shown in this Letter could be 
of great interest to the vast ongoing research on driven 
superconducting qubits \cite{bib:valenzuela}, 
usually modelled also by hamiltonian (\ref{ec:hami}). 

\begin{acknowledgments}

We acknowledge support from CONICET, UBACyT, and ANPCyT (Argentina), the Ministry of Economy and Competitiveness (Spain) under Grants No.~MTM2012-39101 and ICMAT Severo Ochoa SEV2011-0087.
Also, this project has been funded with support from the European Commission; in this respect this publication reflects the views only of the authors, and the Commission cannot be held responsible for any use which may be made of the information contained therein.

\end{acknowledgments}


\begin{thebibliography}{20}
\bibitem{bib:hbarra} Note that we set $\hbar=1$ throughout the Letter.
\bibitem{bib:aclaracion} Note that the AC in model (\ref{ec:hami}) is symmetrical; 
 in the general case, the slopes of the diabatic branches must be taken into account. This leads to the rescaling $A\rightarrow A'=\frac{\Delta \alpha}{2} A$, where $\Delta \alpha$ is the difference of slopes between the branches (in absolute value).
\bibitem{bib:otromet} Note that preparing the system in state $\Ket{LiNC}$ can be done by exciting the molecule at zero field and then adiabatically tuning the field from $\mathcal{E}=0$ to $\mathcal{E}=\mathcal{E}_{\mathrm{dc}}$, as explained in Refs. \cite{bib:murgida,bib:poggi}.
\end{thebibliography}
\end{document}